# Discovering seminal works with marker papers

Robin Haunschild*+ and Werner Marx*

+ corresponding author

* Max Planck Institute for Solid State Research

Heisenbergstr. 1,

70569 Stuttgart

Germany

r.haunschild@fkf.mpg.de, w.marx@fkf.mpg.de

**Abstract**

Bibliometric information retrieval in databases can employ different strategies. Commonly, queries are performed by searching in title, abstract and/or author keywords (author vocabulary). More advanced queries employ database keywords to search in a controlled vocabulary. Queries based on search terms can be augmented with their citing papers if a research field cannot be curtailed by the search query alone. Here, we present another strategy to discover the most important papers of a research field. A marker paper is used to reveal the most important works for the relevant community. All papers co-cited with the marker paper are analyzed using reference publication year spectroscopy (RPYS). For demonstration of the marker paper approach, density functional theory (DFT) is used as a research field. Comparisons between a prior RPYS on a publication set compiled using a keyword-based search in a controlled vocabulary and three different co-citation RPYS (RPYS-CO) analyses show very similar results. Similarities and differences are discussed.

# 1    Introduction

Information retrieval in databases can be performed using different routes. Commonly, searches are performed via search terms (author vocabulary) in the full-text or in certain sections of a paper (e. g., title, abstract, and/or author keywords). Some databases also offer controlled vocabulary (i. e., keywords assigned by the database producer) to be searched. Searches in author vocabulary often require a strategy which is called "interactive query formulation" and was extensively discussed by Wacholder (2011). This strategy was applied for example in Haunschild, Bornmann, and Marx (2016) and Wang, Pan, Ke et al. (2014) to analyze the literature about climate change. A search in controlled vocabulary often needs less search terms and less complicated queries. For example, Haunschild, Barth, and Marx (2016) used a rather concise search query in the controlled vocabulary of $CAplus^{SM}$ to analyze the literature about density functional theory (DFT), a widely used method in the field of computational chemistry.

Besides keyword searches, the citing papers of one specific marker paper (or a few marker papers) can be used to retrieve fundamental literature, see e. g., Marx, Haunschild, and Bornmann (2017). This enables bibliometricians to cover publication sets which are hard to narrow down using keyword searches only.

Our research questions can be formulated as follows: Do citing papers of marker papers represent the historical evolution of research fields in a similar way as publication sets gathered by keyword searches? How can the methodology be used if no suitable marker paper is known? How does the marker paper approach depend on the employed databases?

CitNetExplorer (van Eck & Waltman, 2014, see also http://www.citnetexplorer.nl), a tool based on Eugene Garfield's work on algorithmic historiography, and the corresponding program HistCite (Garfield, 2009; the program is no longer in active development or officially supported) show the time evolution of a given research topic via the citation network of major papers, which have been selected before using other methods. For example, Waaijer and Palmblad (2015) analyzed the historical evolution of the field of mass spectrometry using CitNetExplorer. Garfield and Pudovkin (2003) analyzed the chronological development of the field of nano-crystals and nano-ceramics.

Here, we apply a methodology using a single marker paper (or a few marker papers) for retrieving the set of most influential publications of a topic and analyzing the historical development of the topic. This methodology (RPYS-CO which is a special case of the reference publication year spectroscopy, RPYS) is based on the co-citation network of publications (Small, 1973). RPYS is a bibliometric method for locating seminal papers and the historical roots in publication sets covering specific research topics or fields (Marx, Bornmann, Barth et al., 2014). The method analyzes the cited references of the papers of the relevant publication set. The references most frequently cited are analyzed in graphical and tabular forms. This provides a more objective answer to the question about seminal papers and historical roots (based on the "wisdom of the crowd"). Individual scientists in the field can answer this question only subjectively. However, many scientists with knowledge in the studied field deliver a broader view which is the basis for the interpretation of the RPYS results.

We will compare the results from our RPYS-CO analysis with the previous RPYS analysis by Haunschild, Barth, et al. (2016) which is based on a keyword search in a controlled vocabulary. The scope of our study is on the RPYS-CO method rather than on the seminal papers of DFT themselves. Previously, the methodology has been applied to the history of the greenhouse effect (Marx, Haunschild, Thor et al., 2017). The references within the citing papers of the marker paper are used in a RPYS analysis. The publication set to be analyzed contains all papers which have been co-cited with the marker paper. In case of a few marker papers, the papers of the publication set are co-cited with at least one of the marker papers.

## 2   Methods

### 2.1   Methodology

We used the CRExplorer (see: http://crexplorer.net) to perform the RPYS analysis. The program can be downloaded for free and a comprehensive handbook explaining all functions is available. We used the CRExplorer script language to process the reference variants. The script in **Listing 1** was used to perform the RPYS analysis. The command `importFile` is used to import all WoS papers in the file `"citing_papers.wos.txt"` which were published between 1988 and 2017. The range of reference publication years (RPYs) is restricted to 1950-1990 in order to analyze the same time frame as reported in Haunschild, Barth, et al. (2016). References are cited according to different journal standards, OCR and other errors can happen. Therefore, references "mutate", e.g., a "0" can become an "8" or vice versa. CR variants which are mutated in such a way are called "equivalent CR variants". Clustering and merging equivalent CR variants is done via the commands `cluster` and `merge`. All CRs which were referenced less than 100 times are removed via the `removeCR` command. The value of 100 should be adjusted to the size of the studied data set in terms of cited references. Finally, the command exportFile is used to write the results (CR file and spectrogram file) in CSV format to files. The R package BibPlots (see: https://cran.r-project.org/web/packages/BibPlots/index.html and https://tinyurl.com/y97bb54z) is used to plot the spectrograms.

```
importFile(file: "citing_papers.wos.txt", type: "WOS", RPY:
[1950, 1990, false], PY: [1988, 2017, false], maxCR: 0)
cluster(threshold: 0.75, volume: true, page: true, DOI: false)
```

```
merge()
removeCR( N_CR: [0, 99])
exportFile(file: "full_rpys_CR.csv", type: "CSV_CR")
exportFile(file: "full_rpys_GRAPH.csv", type: "CSV_GRAPH")
```

**Listing 1**: CRExplorer script to perform RPYS-CO analyses

### 2.2    Datasets used

A suitable marker paper should fulfill at least two requirements: (i) it should be cited fairly well considering the topic under study, and (ii) it should reasonably represent the studied topic. Very good candidates would be, e. g., Becke (1988), Kohn and Sham (1965), Hohenberg and Kohn (1964), Lee et al. (1988), Perdew (1986), Perdew, Burke, and Ernzerhof (1996), and Perdew, Ernzerhof, and Burke (1996). The proper choice of suitable marker papers requires at least some knowledge of the topic under study.

Database for marker papers published since 1980: Bibliographic and citation data for papers published since 1980 are available in the Web of Science (WoS, Clarivate Analytics) custom data of our in-house database derived from the Science Citation Index Expanded (SCI-E), Social Sciences Citation Index (SSCI), and Arts and Humanities Citation Index (AHCI) produced by Clarivate Analytics (Philadelphia, USA).

Database for marker papers published before 1980: Bibliographic and citation data for papers published since 1800 are available in the Microsoft Academic (MA) custom data (Sinha, Shen, Song et al., 2015), see also https://aka.ms/msracad, of our in-house database.

A good marker paper should be of high relevance for the field under study. As a first marker paper, we selected the publication by Becke (1988) in which he proposed a very popular density functional approximation for the exchange energy which was for example used together with the LYP correlation functional (Lee, Yang, & Parr, 1988) and in the very popular B3LYP functional (Stephens, Devlin, Chabalowski et al., 1994). Therefore, Becke (1988)

(also known as "Becke88") seems to be a very promising candidate for a marker paper. We exported all papers (n = 34,437) from our in-house database which cited this marker paper.

As a second marker paper, we selected the publication by Kohn and Sham (1965) in which they laid out the basic framework for practical DFT calculations. We exported all publications (n = 23,094), which cited Kohn and Sham (1965), with cited references.

Using a third marker paper, we would like to present the possibility to start with a "bad" marker paper when no "good" marker paper is known. A "good" marker paper should be fairly well cited and should be representable for the topic of interest. We use the publication by Sun, Haunschild, Xiao et al. (2013) as an example of a "bad" marker paper. We exported all publications (n = 69), which cited Sun et al. (2013), with cited references. This "bad" marker paper represents only a rather small part of the topic of interest and has not been cited often enough to be considered as a "good" marker paper.

Unfortunately, the WoS data cannot be shared, but the MA data are available under an open data license (ODC-BY). Interested readers can use our MA data set from https://ivs.fkf.mpg.de/DFT-RPYS/pids_citing_Kohn-Sham-65.wos.zip (DOI 10.5281/zenodo.3579134). We also will be happy to share the full RPYS results of all of the analyses with interested readers.

## 3    Results

### 3.1    RPYS-CO with suitable marker papers

.In the following two subsections, two different marker papers, Becke (1988) and Kohn and Sham (1965), are used.

### 3.1.1 Using Becke (1988) as a marker paper

The paper by Becke (1988) is highly cited. Furthermore, Becke (1988) presents a very popular approximation for the exchange energy functional. Every researcher using this approximation should cite this paper. Therefore, this paper presents a very good candidate for a marker paper. **Fig. 1** shows the number of cited reference (NCR) curves for the RPYS-CO using Becke (1988) as a marker paper and the RPYS from Haunschild, Barth, et al. (2016) for the time frame 1950-1990. The NCR curves indicate how many cited references were published in a specific reference publication year (RPY). Both curves show differences and similarities. **Fig. 2** shows the spectrogram of the RPYS-CO analysis using Becke (1988) as a marker paper. The NCR curves show peaks in some RPYs. Very frequently referenced works can be expected below the peaks. However, usually, many cited references are responsible for a single peak. The peaks are positioned in or around the same RPYs (1951, 1955, 1964/65, 1970, 1972/73/74, 1976/77, 1980, 1985/86, and 1988) but the peak heights differ. The peak papers from the RPYS analysis were discussed in Haunschild, Barth, et al. (2016). The peak papers of the RPYS-CO analysis are listed in **Table 1**.

The CRs 11, 12, 13, 15, and 16 appear in the RPYS-CO but were not mentioned in the RPYS analysis of Haunschild, Barth, et al. (2016). These five CRs of course occurred in the RPYS analysis, too, but did not seem to be as significant as in the RPYS-CO analysis performed in this study. The other 14 CRs of the RPYS-CO also appeared in the RPYS of Haunschild, Barth, et al. (2016). Some CRs even have very similar NCR values, e. g., CR1 with NCR = 793 in the RPYS-CO and NCR = 737 in the RPYS of Haunschild, Barth, et al. (2016). The largest absolute deviation between the results of RPYS and RPYS-CO are found for the marker paper CR18 with NCR = 33,850 in the RPYS-CO and NCR = 14,150 in the RPYS. The peak in the RPY 1976/77 in this RPYS-CO is broader than in the RPYS of Haunschild, Barth, et al. (2016). The different focus can be seen by the comparison of the

NCR values of CR10: NCR = 407 in RPYS-CO and NCR = 6506 in RPYS. Monkhorst and Pack proposed a new method to generate special points in the Brillouin zone which enables more efficient integrations of periodic functions. This method had much more impact in the overall DFT community than in the publication set of our RPYS-CO.

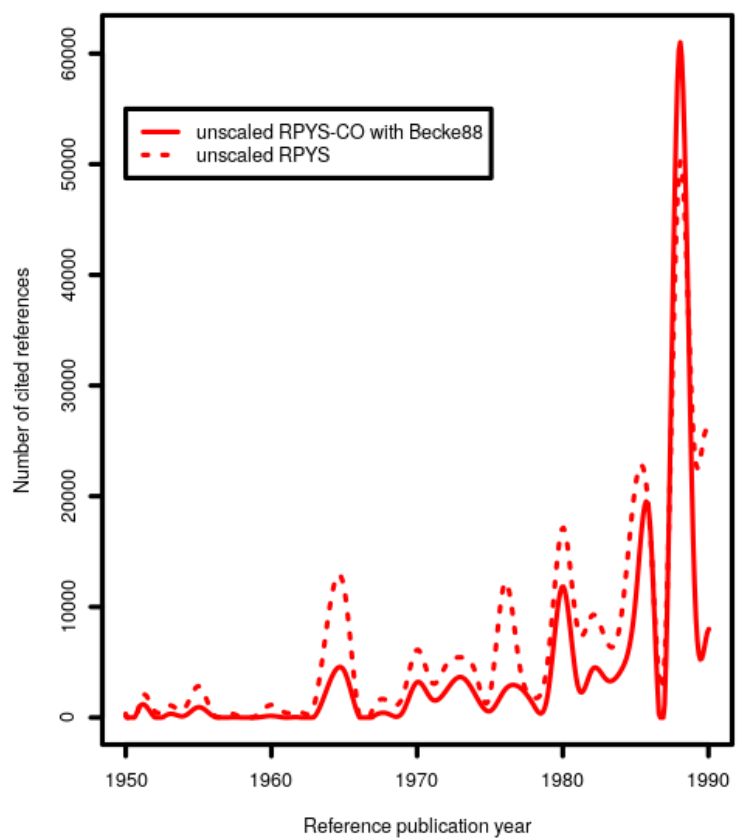

**Fig. 1.** Comparison of NCR curves from the RPYS analysis using DFT papers from a keyword search in controlled vocabulary of the CAS thesaurus for the time frame 1950-1990 from Haunschild, Barth, et al. (2016) with the RPYS-CO analysis in this study using Becke (1988) as a marker paper

In CR11, Ziegler and Rauk proposed a methodology for calculating bonding energies and bond distances using the Hartree-Fock-Slater method. Optimized basis sets for $3d$ orbitals were presented by Hay in CR12. Hirshfeld proposed a molecular partial charge analysis in

CR 13. Hay presented very frequently used ab-initio effective core potentials for molecular calculations in CRs 15 and 16. These CRs had more impact in the publication set of our RPYS-CO than in the RPYS analysis based on keywords as presented by Haunschild, Barth, et al. (2016) although they also appeared in the keyword-based analysis with a relatively lower NCR.

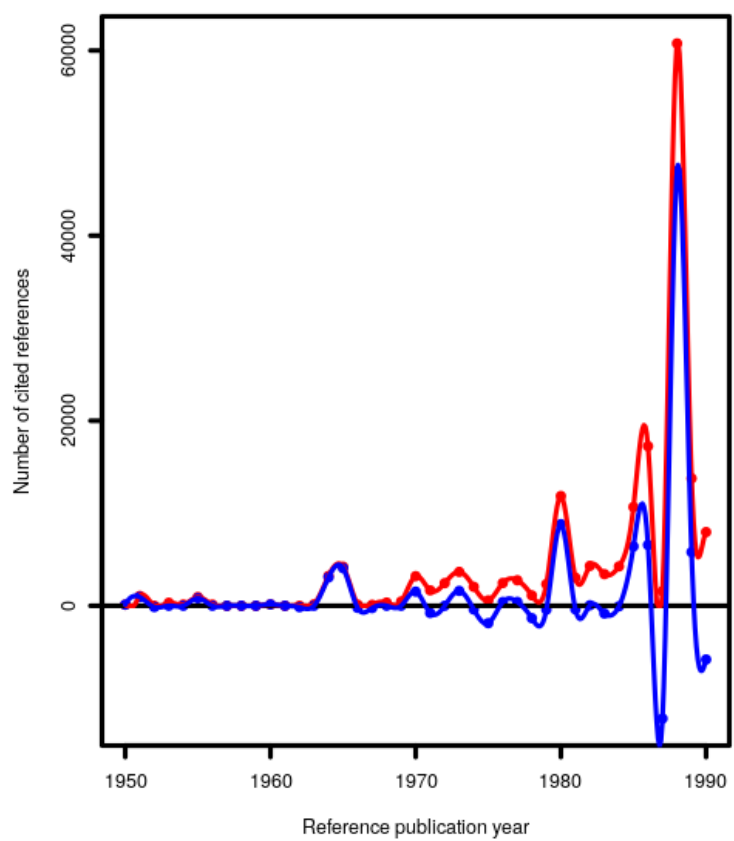

**Fig. 2.** RPYS-CO analysis using papers co-cited with Becke (1988) for the time frame 1950-1990. The red curve and dots show the NCR values. The blue curve and dots show the five-year median deviation. Both curves are used to locate peaks.

**Table 1.** Peak papers of the RPYS-CO using papers co-cited with Becke (1988) for the time frame 1950-1990

| No | RPY | CR | NCR |
|---|---|---|---|
| CR1 | 1951 | Slater JC, 1951, Physical Review, V81, P385 | 793 |
| CR2 | 1951 | Roothaan CCJ, 1951, Reviews of Modern Physics, V23, P69 | 267 |
| CR3 | 1955 | Mulliken RS, 1955, Journal of Chemical Physics, V23, P1833 | 642 |
| CR4 | 1964 | Hohenberg P, 1964, Physical Review B, V136, Pb864 | 2,713 |
| CR5 | 1965 | Kohn W, 1965, Physical Review, V140, P1133 | 3,688 |
| CR6 | 1970 | Boys SF, 1970, Molecular Physics, V19, P553 | 1,584 |
| CR7 | 1972 | Hehre WJ, 1972, Journal of Chemical Physics, V56, P2257 | 1,815 |
| CR8 | 1973 | Harihara PC, 1973, Theoretica Chimica Acta, V28, P213 | 1,957 |
| CR9 | 1973 | Baerends EJ, 1973, Chemical Physics, V2, P41 | 1,446 |
| CR10 | 1976 | Monkhorst HJ, 1976, Physical Review B, V13, P5188 | 407 |
| CR11 | 1977 | Ziegler T, 1977, Theoretica Chimica Acta, V46, P1 | 645 |
| CR12 | 1977 | Hay PJ, 1977, Journal of Chemical Physics, V66, P4377 | 428 |
| CR13 | 1977 | Hirshfeld FL, 1977, Theoretica Chimica Acta, V44, P129 | 398 |
| CR14 | 1980 | Vosko SH, 1980, Canadian Journal of Physics, V58, P1200 | 6,962 |
| CR15 | 1985 | Hay PJ, 1985, Journal of Chemical Physics, V82, P299 | 2,340 |
| CR16 | 1985 | Hay PJ, 1985, Journal of Chemical Physics, V82, P270 | 1,710 |
| CR17 | 1986 | Perdew JP, 1986, Physical Review B, V33, P8822 | 10,308 |
| CR18 | 1988 | Becke AD, 1988, Physical Review A, V38, P3098 | 33,850 |
| CR19 | 1988 | Lee CT, 1988, Physical Review B, V37, P785 | 21,887 |

In fact, we captured the most important seminal papers in **Table 1** as we can see from ordering the CRs by the NCR value. All 10 most frequently occurring CRs appear in **Table 1** except two of them (Dunning (1989) with NCR = 2658 and Parr and Yang (1989) with NCR = 2263). Dunning (1989) proposed very popular atom-centered basis sets. Parr and Yang (1989) is a very popular textbook about DFT. Both CRs were published in 1989. We see that 1989 is on the lower end of the downward slope of the 1988 peak. It is a matter of choice of the scope of the analysis if such RPYs should also be investigated. However, inspection of the most frequently occurring CRs is always recommended. Studies which have a specific topic as a focus, should investigate the RPYS results more deeply than performed here. For example, the CRExplorer also offers advanced indicators to discover papers with significant impact over many citing years (Thor, Bornmann, Marx et al., 2018).

### 3.1.2 Using Kohn and Sham (1965) as a marker paper

The paper Kohn and Sham (1965) is highly cited. Furthermore, the authors laid out the basic framework for practical DFT calculations. Although many researchers in DFT consider the pioneering work of Kohn and Sham (1965) to be textbook knowledge, the paper received 1800 citations in 2017. **Fig. 3** shows the number of cited reference (NCR) curves for the RPYS-CO using Kohn and Sham (1965) as a marker paper and the RPYS from Haunschild, Barth, et al. (2016) for the time frame 1950-1990. The NCR curves show differences and similarities. **Fig. 4** shows the spectrogram of the RPYS-CO analysis using Kohn and Sham (1965) as a marker paper. The peaks are positioned in or around the same RPYs (1951, 1953/54/55, 1964/65, 1972/73, 1976, 1980, 1985/86, and 1988/89) but the peak heights differ for most peaks. The peak papers of the RPYS-CO analysis are listed in **Table 2**. Eleven of the seventeen CRs in **Table 2** also occurred as peak papers in the RPYS-CO analysis using Becke (1988) as a marker paper in the previous subsection. The other six CRs occurred in the previous RPYS-CO analysis and the RPYS analysis by Haunschild, Barth, et al. (2016), too, although not as pronounced peak papers.

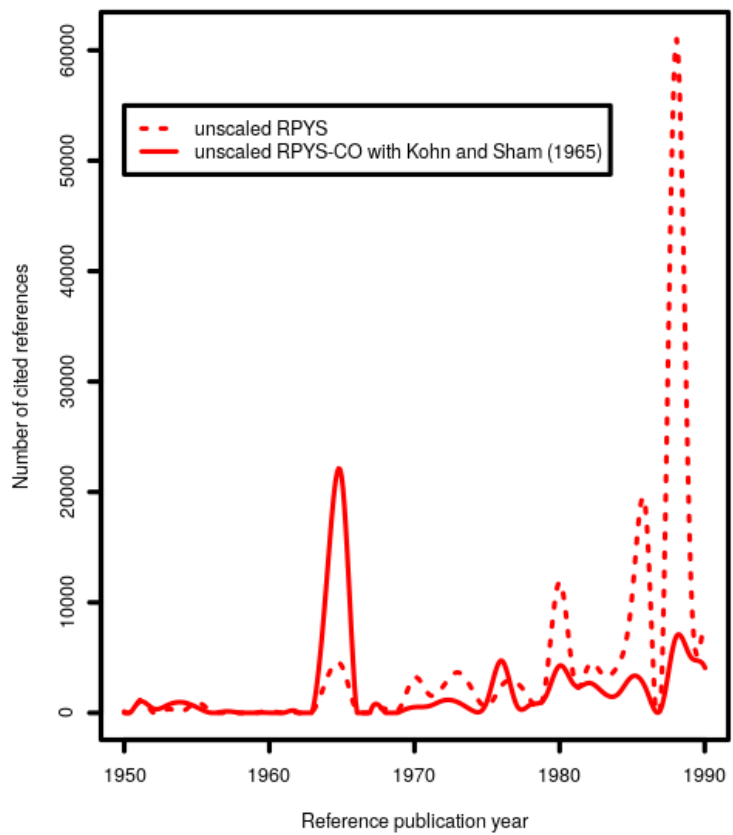

**Fig. 3.** Comparison of NCR curves from the RPYS analysis using DFT papers from a keyword search in controlled vocabulary of the CAS thesaurus for the time frame 1950-1990 from Haunschild, Barth, et al. (2016) with the RPYS-CO analysis in this study using Kohn and Sham (1965) as a marker paper

CR21 is a popular introductory textbook into solid state physics authored by Charles Kittel. CR22 discusses relations between the elastic and plastic properties of pure polycrystalline metals. This CR is important for several applications of DFT to solid state physics. CR26 presents a local exchange-correlation potential for spin-polarized cases. Thereby, this CR extends the work of Hohenberg and Kohn (1964) and Kohn and Sham (1965) to open-shell and broken-symmetry cases. The results in CR30 were used to construct correlation function-als. In CR32, a very popular employed ansatz for molecular dynamics in DFT is proposed. CR36 is a popular textbook about DFT authored by Robert G. Parr and Weitao Yang.

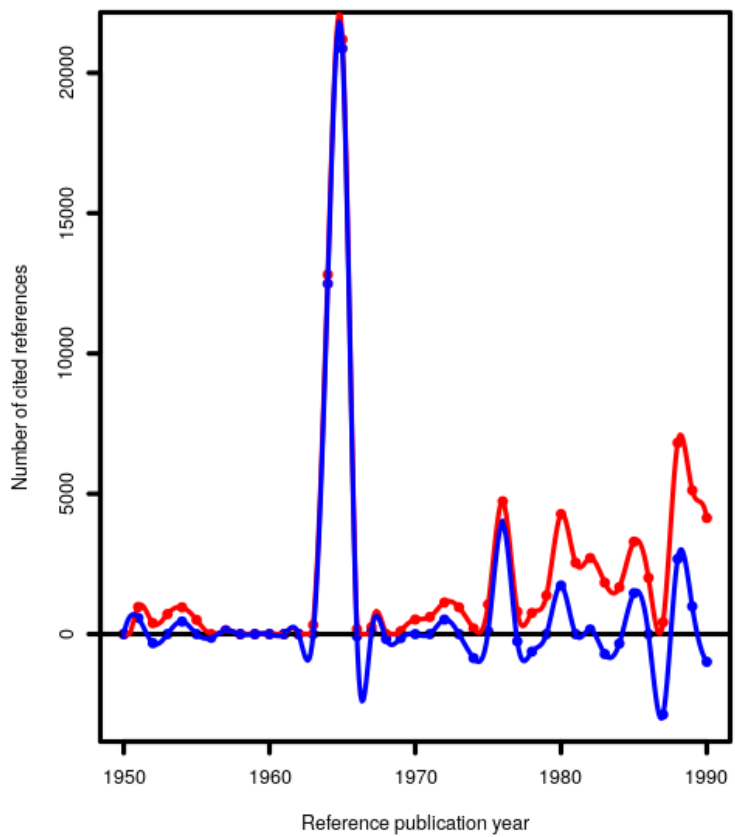

**Fig. 4.** RPYS-CO analysis using papers co-cited with Kohn and Sham (1965) for the time frame 1950-1990. The red curve and dots show the NCR values. The blue curve and dots show the five-year median deviation. Both curves are used to locate peaks.

**Table 2.** Peak papers of the RPYS-CO using papers co-cited with Kohn and Sham (1965) for the time frame 1950-1990

| No | RPY | CR | NCR |
|---|---|---|---|
| CR20 | 1951 | Slater JC, 1951, Physical Review, V81, P385 | 621 |
| CR21 | 1953 | Kittel C, 1953, Introduction to solid state physics | 403 |
| CR22 | 1954 | Pugh SF, 1954, Philosophical Magazine Series 1, V45, P823 | 381 |
| CR23 | 1955 | Mulliken RS, 1955, Journal of Chemical Physics, V23, P1833 | 370 |
| CR24 | 1964 | Hohenberg P, 1964, Physical Review B, V136, Pb864 | 12,700 |
| CR25 | 1965 | Kohn W, 1965, Physical Review, V140, P1133 | 20,455 |
| CR26 | 1972 | von Barth U, 1972, Journal of Physics C: Solid State Physics, V5, P1629 | 524 |
| CR27 | 1972 | Hehre WJ, 1972, Journal of Chemical Physics, V56, P2257 | 364 |
| CR28 | 1973 | Harihara PC, 1973, Theoretica Chimica Acta, V28, P213 | 317 |
| CR29 | 1976 | Monkhorst HJ, 1976, Physical Review B, V13, P5188 | 3,627 |
| CR30 | 1980 | Ceperley DM, 1980, Physical Review Letters, V45, P566 | 1,729 |
| CR31 | 1980 | Vosko SH, 1980, Canadian Journal of Physics, V58, P1200 | 1,492 |
| CR32 | 1985 | Car R, 1985, Physical Review Letters, V55, P2471 | 959 |
| CR33 | 1986 | Perdew JP, 1986, Physical Review B, V33, P8822 | 782 |
| CR34 | 1988 | Lee CT, 1988, Physical Review B, V37, P785 | 3,362 |
| CR35 | 1988 | Becke AD, 1988, Physical Review A, V38, P3098 | 2,429 |
| CR36 | 1989 | Parr RG, 1989, Density-functional theory of atoms and molecules | 1,643 |

### 3.2 RPYS-CO without a suitable marker paper

In order to choose a suitable marker paper, one needs at least some insight into the topic under study. Furthermore, a preliminary query using search terms is helpful for determining the usual citation rate of the topic. In this section, we demonstrate, by applying the RPYS-CO methodology iteratively, the procedure starting with a rather poor marker paper. We choose to start with Sun et al. (2013). This paper has been cited 69 times (date of search 05 March, 2019). For the size of a topic like DFT, even a rather poor marker paper should not be cited much less. This paper is a rather special paper which presents density functional approximations which have not yet been widely applied.

**Listing 1** (without the command "`removeCR`" and the command "`RPY: [1950, 1990, false]`" replaced as "`RPY: [1950, 2017, false]`" in order to also capture newer papers in the initial step) is used for the initial RPYS-CO using Sun et al. (2013) as a marker paper. In the first step, we only look at the ten most frequently occurring CRs ordered by NCR as shown in **Table 3**.

**Table 3.** Ten most frequently occurring CRs of the RPYS-CO using papers co-cited with Sun et al. (2013) for the time frame 1950-1990

| No | RPY | CR | NCR |
| --- | --- | --- | --- |
| CR37 | 2013 | Sun JW, 2013, Journal of Chemical Physics, V138 | 51 |
| CR38 | 1996 | Perdew JP, 1996, Physical Review Letters, V77, P3865 | 45 |
| CR39 | 2003 | Tao JM, 2003, Physical Review Letters, V91 | 37 |
| CR40 | 1965 | Kohn W, 1965, Physical Review, V140, P1133 | 36 |
| CR41 | 2006 | Zhao Y, 2006, Journal of Chemical Physics, V125 | 31 |
| CR42 | 2009 | Perdew JP, 2009, Physical Review Letters, V103 | 29 |
| CR43 | 2012 | Sun JW, 2012, J Chem Phys, V137 | 27 |
| CR44 | 1988 | Becke AD, 1988, Physical Review A, V38, P3098 | 26 |
| CR45 | 2008 | Zhao Y, 2008, Theoretica Chimica Acta, V120, P215 | 25 |
| CR46 | 2008 | Perdew JP, 2008, Physical Review Letters, V100 | 25 |

We see that CR38, CR40, and CR44 were mentioned in the previous section as possible suitable marker papers. Furthermore, CR38 has a rather similar NCR value as our rather poor marker paper (CR37). This is already an indication that our choice of the initial marker paper might not have been very good. Therefore, we use CR38 as a new marker paper in the next step of the iterative RPYS-CO, this time using again **Listing 1**. In the following, CR38 will also be interchangeably referred to as "Perdew (1996)". The resulting NCR curve is compared with the one from the RPYS by Haunschild, Barth, et al. (2016) based on a keyword search in controlled vocabulary in **Fig. 5**. Both NCR curves show peaks at the same locations although the heights of the peaks differ substantially. The RPYS-CO spectrogram using CR38 as a marker paper is shown in **Fig. 6**. The corresponding peak papers are listed in **Table 4**. Nine out of 14 CRs in **Table 4** also appeared as peak papers in the RPYS-CO analysis using Becke (1988) as a marker paper. Additional three CRs (CR48, CR57, and CR58) in **Table 4** also appeared as peak papers in **Table 2** from the RPYS-CO using Kohn and Sham (1965) as a marker paper. The other two CRs also appeared in the other RPYS analyses although not as pronounced peak papers. CR47 studied elastic behavior of a crystalline aggregate. This CR is important for several applications of DFT to solid state physics. CR54 presents studies of electrochemical photolysis of water at a semiconductor electrode. The latter CR is an experimental study which was extensively referenced in DFT papers. The slight differences in the RPYS-CO analyses presented here show the different foci which can be carried over from different maker papers into the RPYS-CO results. At least when studying large topics, it might be advisable to perform multiple iterative RPYS-CO analyses in practice and combine the results.

**Table 4.** Peak papers of the RPYS-CO using papers co-cited with CR38 for the time frame 1950-1990

| No | RPY | CR | NCR |
|---|---|---|---|
| CR47 | 1952 | Hill R, 1952, Proceedings of the Physical Society of London Section A, V65, P349 | 1185 |
| CR48 | 1954 | Pugh SF, 1954, Philosophical Magazine, V45, P823 | 1294 |
| CR49 | 1955 | Mulliken RS, 1955, Journal of Chemical Physics, V23, P1833 | 833 |
| CR50 | 1964 | Hohenberg P, 1964, Physical Review B, V136, PB864 | 7509 |
| CR51 | 1965 | Kohn W, 1965, Physical Review, V140, P1133 | 8946 |
| CR52 | 1970 | Boys SF, 1970, Molecular Physics, V19, P553 | 1138 |
| CR53 | 1972 | Hehre WJ, 1972, Journal of Chemical Physics, V56, P2257 | 628 |
| CR54 | 1972 | Fujishima A, 1972, Nature, V238, P37 | 605 |
| CR55 | 1976 | Monkhorst HJ, 1976, Physical Review B, V13, P5188 | 13558 |
| CR56 | 1980 | Vosko SH, 1980, Canadian Journal of Physics, V58, P1200 | 2180 |

| CR57 | 1980 | Ceperley DM, 1980, Physical Review Letters, V45, P566 | 1980 |
| CR58 | 1985 | Car R, 1985, Physical Review Letters, V55, P2471 | 1242 |
| CR59 | 1988 | Lee CT, 1988, Physical Review B, V37, P785 | 4981 |
| CR60 | 1988 | Becke AD, 1988, Physical Review A, V38, P3098 | 4048 |

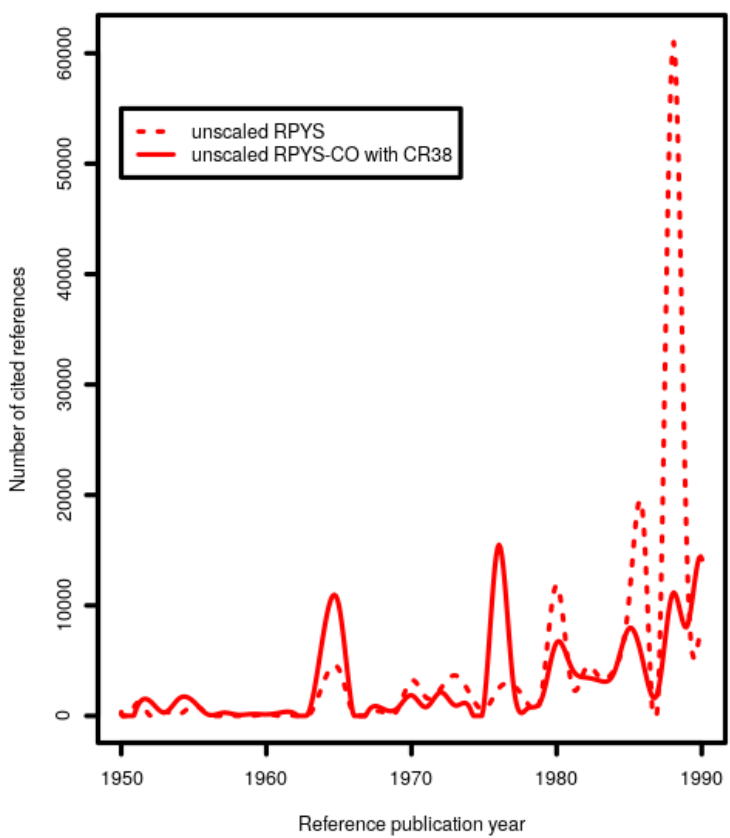

**Fig. 5.** Comparison of NCR curves from the RPYS analysis using DFT papers from a keyword search in controlled vocabulary of the CAS thesaurus for the time frame 1950-1990 from Haunschild, Barth, et al. (2016) with the RPYS-CO analysis in this study using CR38 as a marker paper

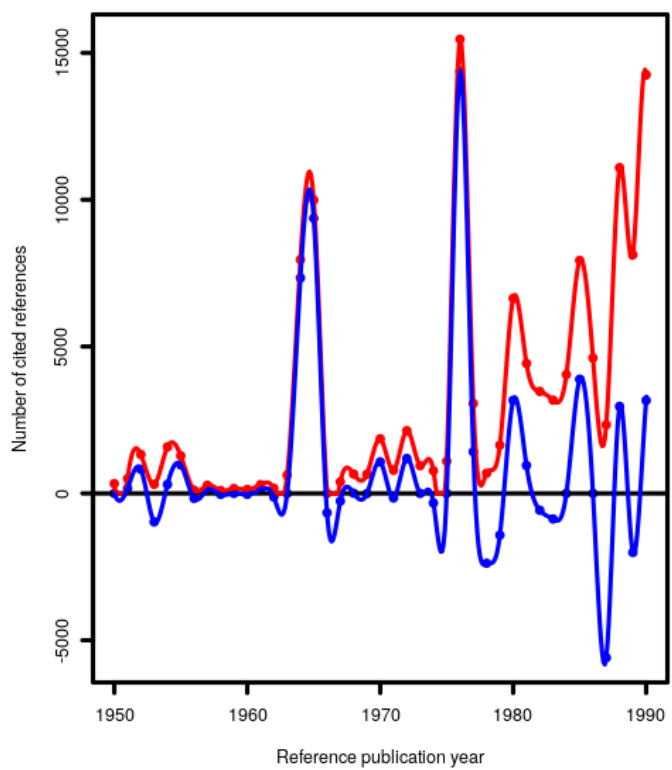

**Fig. 6.** RPYS-CO analysis using papers co-cited with CR38 for the time frame 1950-1990. The red curve and dots show the NCR values. The blue curve and dots show the five-year median deviation. Both curves are used to locate peaks.

### 3.3 Comparison of the four different RPYS results

**Fig. 7** shows the NCR curves of the three different RPYS-CO analyses in this study in comparison to the RPYS analysis using DFT papers from a keyword search in controlled vocabulary of the CAS thesaurus for the time frame 1950-1990 from Haunschild, Barth, et al. (2016). Most peaks are positioned at the same RPYs. As the tables in the previous subsections show, most RPYS-CO analyses produce also the same peak papers. Due to a different focus of each RPYS-CO analysis, different peak heights occur, and in some RPYs different peak papers are found. The most significant differences between the RPYS analyses in peak locations and peak widths can be observed in the early 1950s and early 1970s.

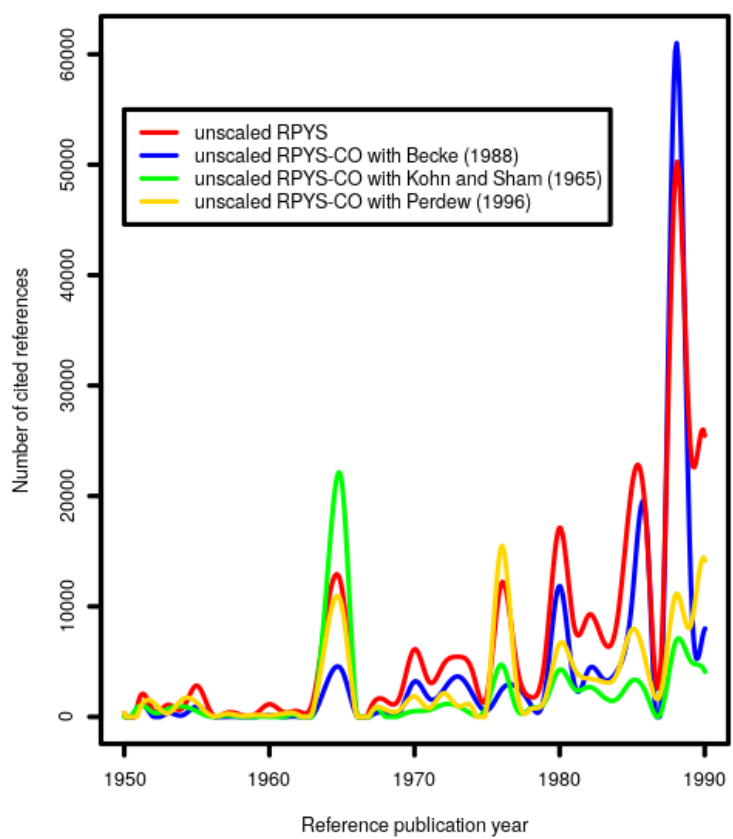

**Fig. 7.** Comparison of NCR curves from the RPYS analysis using DFT papers from a keyword search in controlled vocabulary of the CAS thesaurus for the time frame 1950-1990 from Haunschild, Barth, et al. (2016) with the three different RPYS-CO analyses in this study

From comparing our RPYS-CO analyses, we can summarize that most peak papers are extracted by all explored variants. The peak papers from one RPYS-CO analysis, which did not occur as peak papers in other RPYS-CO analyses, still occurred as very frequently referenced publications. Overall, the different RPYS-CO analyses extracted the same landmark papers, only weighted with different importance. This different importance originates from the different choices of marker papers. The choice of marker papers determines the spe-cial focus within a certain field. We expect the RPYS analysis based on the keyword search in the CAplus database (Haunschild, Barth, et al., 2016) to provide the most realistic perspective because it is based on intellectual indexing of CAS. All of the RPYS-CO analyses agree with the RPYS analysis to a surprisingly high extent.

## 4   Discussion and Conclusions

Overall, the results of the RPYS-CO analyses presented here and the RPYS of Haunschild, Barth, et al. (2016) are very similar although the methodology and the employed database are quite different. Haunschild, Barth, et al. (2016) started from a keyword search in index terms of the CAplus database (controlled vocabulary of the database provider) while the RPYS-CO analyses performed in this study are based on papers co-cited with one marker paper in a bibliometric database (here WoS or MA). Despite the different approaches and different databases, quite similar results were obtained.

We summarize shortly by answering the research questions posed in the introduction of this paper. It is a viable approach to define a research field by the publications co-cited with a suitable marker paper. Even if a suitable marker paper is not known, it is possible to iterate from a reasonable initial paper towards a suitable marker paper. The dependency on the employed databases seems to be low as long as the topic of interest is covered well in the database of choice. There seems to be a more significant dependency on the choice of the marker paper, because this determines the search results as much as the construction of a search query in a keyword-based search.

The approach of using a marker paper for finding other seminal papers in research fields might become an interesting tool for scientists to explore their research fields in addi-tion to a keyword-based literature search. If a good marker paper is not known a priori, the RPYS-CO methodology can be applied iteratively.

The RPYS-CO analysis has several advantages over built-in functionalities of several databases: (i) not only source records of the database can be found but also seminal papers which appear only in the cited references. (ii) The CRExplorer provides additional analysis features, such as filtering for papers which had a significant impact over many citing years by using the advanced indicators. (iii) The RPYS-CO methodology is not restricted to a specific database. In principle, the RPYS-CO methodology can be applied to datasets from any database which includes cited references.

The focus on the cited references, however, has a disadvantage: Search results have to be processed outside the database or reimported into the database. Such a reimport is usually not complete as non-source records appear in the results of an RPYS analysis.

In contrast to using previously employed tools (such as CitNetExplorer and HistCite, for smaller publication sets) to show the time evolution of research topics, using CRExplorer and applying the RPYS-CO method aims at detecting the publications of most importance for the relevant community during the evolution of a given research topic. An alternative method for retrieving relevant literature based on co-citations is the related records search function offered by, e. g., WoS and MA. However, this method retrieves a publication set without any weighting with regard to the citation impact within the relevant community.

We showed that RPYS-CO works using different databases. Future work should apply the method to other research topics. The RPYS-CO methodology is most appropriate in cases where it is very hard (or even impossible) to encompass a scientific field by search-term- or keyword-based searches. For example, Scheidsteger and Haunschild (2019) have used RPYS-CO for analyzing the early history of the field of solar energy meteorology, because it turned out to be very difficult to define the publication set via keyword- or search-term-based queries. Similar problems can arise due to ambiguity of keywords used for the search query (e.g., tea finds not only literature related to beverage tea but also literature related to triethylamine which is often abbreviated as TEA. In such cases, the RPYS-CO methodology as proposed here can be very helpful because it does not rely on a topical- or keyword-based search query.

# Acknowledgements

The bibliometric data used in this paper are from Max Planck Society's in-house databases. The WoS in-house database is developed and maintained in cooperation with the Max Planck Digital Library (MPDL, Munich). It is derived from the Science Citation Index Expanded (SCI-E), Social Sciences Citation Index (SSCI), Arts and Humanities Citation Index (AHCI) prepared by Clarivate Analytics, formerly the IP & Science business of Thomson Reuters (Philadelphia, Pennsylvania, USA). The MA in-house database is a locally maintained database at the Max Planck Institute for Solid State Research derived from the Microsoft Academic database.